\documentstyle[preprint,aps]{revtex}
\newcommand{\be}{\begin{equation}}
\newcommand{\ee}{\end{equation}}
\begin{document}
\title{Vortex Velocities in the $O(n)$ Symmetric TDGL Model}
\author{Gene F. Mazenko}
\address{The James Franck Institute and Department of Physics}
\address{The University of Chicago, Chicago, Illinois 60637}
\date{\today}
\maketitle
 
\begin{abstract}

An explicit expression for the vortex velocity field as a function of the
order parameter field is derived for the case of point defects in the
$O(n)$ symmetric time-dependent Ginzburg-Landau model.  
This expression is used to find the 
vortex velocity probability distribution in the gaussian closure
approximation in the case of phase ordering kinetics for a nonconserved
order parameter.  The velocity
scales as $L^{-1}$ in scaling regime where $L\approx t^{1/2}$ and
t is the time after the quench.

\end{abstract}

The importance of the role of defects in understanding a variety of problems in 
physics is clear.  In certain cosmological\cite{1} 
and phase 
ordering \cite{2}
problems  key questions involve an understanding of the  evolution and 
correlation among defects like vortices, monopoles,  disclinations, etc.  
In studying such objects in a field theory  questions 
arises as to how one can define quantities like the density of vortices 
and an 
associated vortex velocity field.  The purpose of this paper is to
identify 
the
appropriate vortex-velocity field in the context of an $O(n)$ 
symmetric
time-dependent Ginzburg-Landau (TDGL) model for the case of point defects 
where
$n=d$ and $d$ is the spatial dimensionality.  Using this rather 
general
definition for the velocity field  the distribution of 
velocities is determined in
the case of the late state phase ordering using the gaussian
closure approximation for a nonconserved order parameter.
The physical results are that the velocity scales as
$L(t)^{-1}$ where $L(t)\approx  t^{1/2}$ is the characteristic 
scaling length for the order parameter correlation function
which grows with time t after the quench.   The 
vortex velocity 
probability distribution function is given in this approximation by
\be
P(\vec{v}_{0})=\frac{\Gamma (\frac{n}{2}+1)}{(\pi \bar{v}^{2})^{n/2}}
\frac{1}{\Biggl( 1+(\vec{v}_{0})^{2}/\bar{v}^{2}\Biggr)^{(n+2)/2}}
\label{eq:1}
\ee
where the parameter $\bar{v}$ is defined below and varies as $L^{-
1}$ for
long times.

The focus here is on the defect dynamics generated by the TDGL 
model satisfied by a nonconserved $n$-component vector order 
parameter $\vec{\psi}(\vec{r},t)$:
\be
\frac{\partial \vec{\psi}}{\partial t}=\vec{K}\equiv
-\Gamma \frac{\delta F}{\delta \vec{\psi}}  +\vec{\eta}
\label{eq:2}
\ee 
where $\Gamma $ is a kinetic coefficient,  $F$ is a Ginzburg-Landau
effective free energy assumed to be of the form
\be
F=\int ~d^{d}r \biggl( \frac{c}{2}(\nabla \vec{\psi})^{2}
+V(|\vec{\psi}|)\biggr)
\ee
where $c > 0$ and the potential is assumed to be of the 
degenerate double-well 
form.
$\vec{\eta }$ is a thermal noise which is related to $\Gamma$ 
by a
fluctuation-dissipation theorem.

Consider a system with $n=d$ where there are
topologically stable point defects\cite{3} formed
in a phase ordering 
system
(quenched for example from a high temperature disordered state to 
a temperature below the order temperature).  As pointed out by 
Halperin\cite{4},
and exploited by Liu and Mazenko\cite{5}, the vortex density for such a 
system can
be written as
\be
\rho =\delta (\vec{\psi}){\cal D}
\ee
where ${\cal D}$ is the Jacobian (determinant) for the change of 
variables from the set of vortex positions $r_{i}(t)$
(where $\vec{\psi}$ vanishes)  to the field
$\vec{\psi}$:
\be
{\cal D}=\frac{1}{n!}\epsilon_{\mu_{1},\mu_{2},...,\mu_{n}}
\epsilon_{\nu_{1},\nu_{2},...,\nu_{n}}
\nabla_{\mu_{1}}\psi_{\nu_{1}}
\nabla_{\mu_{2}}\psi_{\nu_{2}}....
\nabla_{\mu_{n}}\psi_{\nu_{n}}
\ee
where $\epsilon_{\mu_{1},\mu_{2},...,\mu_{n}}$ is the 
$n$-dimensional fully anti-symmetric tensor and
summation over repeated indices here and below is implied.

The first goal here is to  derive the equation of motion 
satisfied
by $\rho$.  Toward this end one needs two identities whose proof is
relatively straightforward.  The first identity is given by:
\be
\frac{\partial {\cal D}}{\partial t}=\nabla_{\alpha}J_{\alpha}^{(K)}      ~~~~~~Identity~ I
\ee
where, for a general vector $\vec{A}$, the current $J_{\alpha}^{(A)}$
is defined as
\be
J_{\alpha}^{(A)}
=\frac{1}{(n-1)!}\epsilon_{\alpha,\mu_{2},...,\mu_{n}}
\epsilon_{\nu_{1},\nu_{2},...,\nu_{n}}
A_{\nu_{1}}
\nabla_{\mu_{2}}\psi_{\nu_{2}}....
\nabla_{\mu_{n}}\psi_{\nu_{n}}~~~.
\ee
Identity I is just a statement that the determinant ${\cal D}$ is a
conserved invariant.  Notice that the superscript
$K$ on $J$ in this identity is defined by the right hand side of
Eq.(\ref{eq:2}).  The second identity takes the form for general 
vector $\vec{A}$:
\be
J_{\alpha}^{(A)}\nabla_{\alpha}\psi_{\beta}=A_{\beta}
{\cal D}~~~~~~~~~~~~Identity ~II
\ee
Identity II, after using the chain-rule for differentiation,
leads directly to the result
\be
{\cal D}\frac{\partial}{\partial t}\delta (\vec{\psi})=
J_{\beta}^{(K)}\nabla_{\beta}\delta (\vec{\psi}) ~~~.
\ee
When this result is combined with Identity I, one easily obtains 
the equation of motion for 
the
vortex density
\be
\frac{\partial\rho}{\partial t}=\nabla_{\beta}\Biggl[\delta 
(\vec{\psi})J_{\beta}
^{(K)}\Biggr] ~~~.
\label{eq:10}
\ee
This continuity equation
reflects the fact that the vortex ${\it charge}$ is conserved.  A 
key point here is that $J_{\beta}^{(K)}$ is multiplied by the vortex 
locating
$\delta$-function.  This means that one can replace $\vec{K}$ 
in $\vec{J}^{(K)}$ by the 
part of
$\vec{K}$ which does not vanish as $\vec{\psi}\rightarrow 0$.  Thus 
in the case of a nonconserved order parameter one can replace 
$J_{\beta}^{(K)}$ in the
continuity equation by
\be
J_{\beta}^{(2)}
=\frac{1}{(n-1)!}\epsilon_{\beta,\mu_{2},...,\mu_{n}}
\epsilon_{\nu_{1},\nu_{2},...,\nu_{n}}
\Biggl[\Gamma c\nabla^{2}\psi_{\nu_{1}}+\eta_{\nu_{1}}\Biggr]
\nabla_{\mu_{2}}\psi_{\nu_{2}}....
\nabla_{\mu_{n}}\psi_{\nu_{n}}~~~.
\ee
In the case of a conserved order parameter the current for $\rho$
is more complicated  because of the overall gradients acting in
$\vec{K}$.
Because of the standard form of the continuity equation Eq.(\ref{eq:10}), 
it is clear that one 
can identify the vortex velocity field as
\be
v_{\alpha}=-\frac{J_{\alpha}^{(2)}}{{\cal D}}
\label{eq:12}
\ee
where it is assumed that the velocity field is used inside expressions
multiplied by the vortex
locating  $\delta$-function.  This is the primary result of this paper.  
It gives
one an explicit expression for the vortex velocity  field in terms of 
the
original order parameter field.  In the particular case of $n=d=2$ one has
the more explicit result
\be
v_{\alpha}=\Gamma c \frac{\epsilon_{\alpha \mu}(\nabla_{\mu}
\psi_{x}\nabla^{2}\psi_{y}-\nabla_{\mu}\psi_{y}\nabla^{2}\psi_{x})}
{\epsilon_{\nu\sigma}\nabla_{\nu}\psi_{x}\nabla_{\sigma}\psi_{y}} ~~~.
\ee
Notice that the result given by Eq.(\ref{eq:10}) does not depend on the
details of the TDGL model, only that the equation of motion is first
order in time.

As an important application of the result Eq.(\ref{eq:12}) for 
$\vec{v}$ consider the 
velocity
probability distribution function defined by
\be
n_{0} P(\vec{v}_{0})\equiv
\langle n\delta (\vec{v}_{0}-\vec{v})\rangle
\ee
where $\vec{v}_{0}$ is a reference velocity, 
$n =\delta (\vec{\psi})|{\cal D}|$ is the unsigned defect density,
and $n_{0}=\langle n\rangle$.
We determine $P$ using the gaussian closure method\cite{5a,6,7,8}
 which has been 
successful in 
determining the scaling function for the order parameter correlation
function.  The first step is to express the order parameter in terms of 
an
auxiliary field $\vec{m}$ which is  assumed, to a first approximation, 
to
have a gaussian distribution.  In the theory developed in Ref.\cite{6},
the relationship between the order 
parameter
and the auxiliary field is given as a solution to the classical interface
equation 
\be
\nabla_{m}^{2}\vec{\psi} (\vec{m}) = V^{\prime}(|\psi 
|)\hat{\psi}
\ee
where the auxilliary field serves as the coordinate labelling the 
distance to the defect nearest to space point $\vec{r}$ at time t.
The solution of this equation for a charge one vortex is of the form
\be
\vec{\psi}(\vec{m})=A(|\vec{m}|)\hat{m}
\ee
where $A(|\vec{m}|)$ vanishes linearly with $m$ for small $m$
with the next term of ${\cal O}(m^{3})$.   It 
is then
easy to show that one can replace $\vec{\psi}$ by $\vec{m}$ in the
expression for $\vec{v}$.  One can determine $P(\vec{v}_{0})$ by 
first
evaluating the more general probability distribution
\be
G(\xi ,\vec{b})=\langle \delta (\vec{m} )
\delta (\xi_{\mu}^{\nu}-\nabla_{\mu}m_{\nu })
\delta (\vec{b}-\nabla^{2}\vec{m})\rangle
\ee
since
\be
n_{0}P(\vec{v}_{0})=
\int d^{n}b \prod_{\mu ,\nu}d\xi_{\mu}^{\nu}
|{\cal D}(\xi )|
\delta (\vec{v}_{0}-\vec{v}(\vec{b},\xi ))G(\xi ,\vec{b})
\ee
where
\be
\vec{v}(\vec{b},\xi ))=-\frac{\vec{J}^{(2)}(\vec{b},\xi) }{{\cal D}(\xi )}
\ee
with
\be
{\cal D}(\xi )=\frac{1}{n!}\epsilon_{\mu_{1},\mu_{2},...,\mu_{n}}
\epsilon_{\nu_{1},\nu_{2},...,\nu_{n}}
\xi_{\mu_{1}}^{\nu_{1}}
\xi_{\mu_{2}}^{\nu_{2}}....
\xi_{\mu_{n}}^{\nu_{n}}
\ee
and
\be
J_{\alpha}^{(2)}(\vec{b},\xi )
=\frac{1}{(n-1)!}\epsilon_{\alpha,\mu_{2},...,\mu_{n}}
\epsilon_{\nu_{1},\nu_{2},...,\nu_{n}}
\Gamma c b_{\nu_{1}}
\xi_{\mu_{2}}^{\nu_{2}}....
\xi_{\mu_{n}}^{\nu_{n}} ~~~.
\ee
In this last expression it has been assumed that the quench is to 
zero temperature\cite{9} so 
that the noise can be set to zero.  The gaussian average determining 
$G(\xi ,\vec{b})$ is 
relatively
straightforward to evaluate with the result:
\be
G(\xi ,\vec{b})=\frac{1}{(2\pi S_{0})^{n/2}}
\frac{e^{-\frac{1}{2\bar{S}_{4}}\vec{b}^{2}}}{(2\pi \bar{S}_{4})^{n/2}}
\frac{1}{(2\pi S^{(2)})^{n^{2}/2}}
exp\biggl[-\frac{1}{2S^{(2)}}\sum_{\mu 
,\nu}(\xi_{\mu}^{\nu})^{2}\biggr]
\ee
where, $S_{0}=\frac{1}{n}\langle \vec{m}^{2}\rangle$ 
is proportional to $L^{2}$,
\be
S^{(2)}=\frac{1}{n^{2}}\langle (\nabla \vec{m})^{2}\rangle
\ee
and
\be
\bar{S}_{4}=\frac{1}{n}\langle (\nabla^{2}\vec{m})^{2}\rangle
-\frac{\biggl(nS^{(2)}\biggr)^{2}}{S_{0}} ~~~.
\ee
The quantities $S_{0}, S^{(2)},\bar{S}_{4}$ are determined from the 
theory
for the order parameter correlation function.
Using this result for $G(\xi ,\vec{b})$ in the expression for the 
probability
distribution one can use the usual integral representation for the 
$\delta$-function to perform the integration over the $\vec{b}$ field 
to
obtain
\be
n_{0}P(\vec{v}_{0})=\int \prod_{\mu ,\nu}d\xi_{\mu}^{\nu}
\frac{|{\cal D}(\xi )|}{(2\pi S^{(2)})^{n^{2}/2}}
exp\biggl[-\frac{1}{2S^{(2)}}\sum_{\mu 
,\nu}(\xi_{\mu}^{\nu})^{2}\biggr]
\frac{1}{(4\pi^{2}S_{0} \gamma )^{n/2}}\frac{1}{\sqrt{det M}}
exp \Biggl[-\frac{1}{2\gamma }\sum_{\mu ,\nu}
v_{0}^{\mu}[M^{-1}]_{\mu ,\nu }v_{0}^{\nu}\Biggr]
\ee
where $\gamma =\bar{S}_{4}(\Gamma c)^{2}$, and
the matrix $M$ is given by
\be
M_{\alpha ,\beta }=\frac{1}{{\cal D}^{2}[(n-
1)!]^{2}}\epsilon_{\alpha,\mu_{2},...,\mu_{n}}
\epsilon_{\nu,\nu_{2},...,\nu_{n}}
\xi_{\mu_{2}}^{\nu_{2}}....\xi_{\mu_{n}}^{\nu_{n}}
\epsilon_{\beta,\mu_{2}',...,\mu_{n}'}
\epsilon_{\nu,\nu_{2}',...,\nu_{n}'}
\xi_{\mu_{2}'}^{\nu_{2}'}....\xi_{\mu_{n}'}^{\nu_{n}'}~~~~.
\ee
It is straightforward to obtain the rather clean results
\be
det (M) =\frac{1}{({\cal D})^{2}}
\ee
and
\be
M^{-1}_{\alpha \beta} 
=\sum_{\nu}\xi_{\alpha}^{\nu}\xi_{\beta}^{\nu}
\ee
so that
\be
n_{0}P(\vec{v}_{0})=\int\prod_{\mu ,\nu}d\xi_{\mu}^{\nu}
\frac{1}{(2\pi S^{(2)})^{n^{2}/2}}
exp\biggl[-\frac{1}{2S^{(2)}}\sum_{\mu 
,\nu}(\xi_{\mu}^{\nu})^{2}\biggr]
\frac{{\cal D}^{2}}{(4\pi^{2}S_{0} \gamma )^{n/2}}
exp \Biggl[-\frac{1}{2\gamma }\sum_{\alpha ,\beta ,\nu}
v_{0}^{\alpha}\xi_{\alpha}^{\nu}\xi_{\beta}^{\nu}v_{0}^{\beta}\Biggr]
~~~.
\ee
The remaining integrals look formidable but can be carried out if one
makes a
transformation from $\xi_{\alpha}^{\nu}$ to $\chi_{\alpha}^{\nu}$ 
via
\be
\xi_{\alpha}^{\nu}=N_{\alpha ,\beta }\chi_{\beta }^{\nu}
\ee
such that 
\be
\frac{1}{S^{(2)}}\sum_{\alpha ,\nu}(\xi_{\alpha }^{\nu})^{2}
+\frac{1}{\gamma }\sum_{\alpha ,\beta ,\nu}
v_{0}^{\alpha}\xi_{\alpha}^{\nu}\xi_{\beta}^{\nu}v_{0}^{\beta}
=\sum_{\alpha ,\nu}(\chi_{\alpha }^{\nu})^{2}~~~~.
\ee
One  easily finds that
\be
N_{\alpha \beta}=\sqrt{S^{(2)}}\Biggl[\delta_{\alpha \beta}
+[\frac{1}{\sqrt{1+(\vec{v}_{0})^{2}/\bar{v}^{2}}}-
1]\hat{v}_{0}^{\alpha}
\hat{v}_{0}^{\beta}\Biggr]
\ee
where
\be
\bar{v}^{2}=\gamma /S^{(2)}=(\Gamma c)^{2}\frac{\bar{S}_{4}}{S^{(2)}}~~~.
\ee
After changing variables from $\xi$ to $\chi$ one obtains
\be
n_{0}P(\vec{v}_{0})=\biggl(\frac{S^{(2)}}{2\pi S_{0}}\biggr)^{n/2}
\biggl(\frac{1}{2\pi\bar{v}^{2}}\biggr)^{n/2}
\frac{\tilde{J}}{[1+(\vec{v}_{0})^{2}/\bar{v}^{2}]^{(n+2)/2}}
\ee
where
$\tilde{J}$ is the remaining dimensionless integral over the
$\chi$'s which can be
be evaluated directly as a separable product of gaussian integrals with
the simple result $\tilde{J}=n!$.
Since $P(v_{0})$ is normalized to one, we find on integration over
$\vec{v}_{0}$ the result
\be
n_{0}=\biggl(\frac{S^{(2)}}{2\pi S_{0}}\biggr)^{n/2}
\frac{n!}{2^{n/2}\Gamma(\frac{n}{2}+1)}
\ee
which agrees with the result found by Liu and Mazenko\cite{5}
using a more indirect method.  After using this result for
$n_{0}$ one
finally
obtains the result given by Eq.(\ref{eq:1}).  This result basically
says that the probability of finding a large velocity decreases
with time.  However, since this distribution falls off only as
$v_{0}^{-(n+2)}$ for large $v_{0}$ only the first moment beyond the
normalization integral exists.  This seems to imply the existance of
a source of large velocities.  It seems likely\cite{14} that this is
associated with vortex-antivortex final annihilation.

The determination of $S_{0}, S^{(2)}$,
$\bar{S}_{4}$ and $\bar{v}$ requires a theory for the auxiliary field 
correlation
function 
\be
C_{0}(12)=\frac{1}{n}\langle \vec{m}(1)\cdot \vec{m}(2)\rangle ~~~~.
\ee
There are two theories available and both are of the gaussian closure
type assumed above.  One, due to Ohta, Jasnow, and
Kawasaki(OJK)\cite{10}, essentially
postulates that  $C_{0}(12)$ is a gaussian
\be
 C_{0}(12)=S_{0}e^{-\vec{r}^{2}/(2L^{2})}
\ee
where $\vec{r}=\vec{r_{1}}-\vec{r}_{2}$.  In this case one easily
finds that 
\be
\bar{v}^{2}=\frac{2d}{L^{2}}(\Gamma c )^{2}
\ee
where the coefficient of $L^{2}/t$ is undetermined in the theory of
OJK.  Using the theory developed in Ref.\cite{11}  for $n=2$
one finds self-consistently\cite{12} that
\be
\bar{v}^{2}=(1+\frac{\pi}{4\mu })\frac{(\Gamma c)^{2}}{t}
\ee
where $\mu = 0.53721..$ is the eigenvalue determined within the 
theory\cite{11}.

One can go forward and extend these ideas to treat two-point 
velocity correlation
functions and string-like defects ($n=d-1$) as will be discussed 
elsewhere.

\acknowledgements
  I thank Professor Alan Bray for useful comments on
this work.

\pagebreak


\begin{references}
 
\bibitem{1} For a recent review of the role of defects in both a
cosmological and condensed matter context see 
{\it Formation~and~Interactions~of~Topological~Defects},
ed by A-C Davis and R. Brandenberger, NATO ASI Series B:
Physics Vol. 349, (Plenum, New York, 1995).
\bibitem{2} A.J. Bray, Adv. Phys. {\bf 43}, 357 (1994).  This is an
excellant recent review of phase order kinetics from one of the leaders
in the field.
\bibitem{3} N.D. Mermin, Rev. Mod. Phys. {\bf 51}, 591 (1979) gives a
good review of topological defects in ordered media.
\bibitem{4} B.I . Halperin, in {\it Physics~of~Defects},
edited by R. Balian et al. (North-Holland, Amsterdam, 1981).
\bibitem{5} F. Liu and G. F. Mazenko, Phys. Rev. B{\bf 46}, 5963 (1992).
\bibitem{5a} G. F. Mazenko, Phys. Rev. B{\bf 42}, 4487 (1990).
\bibitem{6}  F. Liu and G.F. Mazenko, Phys. Rev. B{\bf 45}, 6989 (1992).
\bibitem{7}  A.J. Bray and K. Humayun, J. Phys. A{\bf 25},2191 (1992).
\bibitem{8}  H. Toyoki,Phys. Rev. B{\bf 45}, 1965 (1992).
\bibitem{9}  Because of the growing length L in the problem
the role of temperature is typically irrelevant as long as
the final temperature T is less than the critical temperature
$T_{c}$.
\bibitem{14}  Professor Alan Bray informs me that he has a scaling
argument supporting the value of large velocity exponent in the probability
distribution.
\bibitem{10} T. Ohta, D. Jasnow, and K.  Kawasaki, Phys. Rev. Lett.
{\bf 49}, 1223 (1982).  See Ref.\cite{2} for extensions of this basic
method.
\bibitem{11}  G.F. Mazenko and R.A. Wickham, Cond-mat/9607152, to be published
Phys Rev E.
\bibitem{12}  The result $\bar{S}_{4}\approx L^{-2}$ is nontrivial and
assumes that there are no correction to scaling terms of the form
$C_{0}=S_{0}-\frac{S^{(2)}}{2}(r^{2}+b_{4}r^{4}+...)$
for small $r$ where $b_{4}$ is a constant as $t\rightarrow\infty$.
A detailed self-consistent analaysis shows that $b_{4}=0$ if $\bar{S}_{4}$ is to
be positive at late times.
\end{references}
\end{document}